\documentclass[preprint]{vgtc}               %

\graphicspath{{figures/}{pictures/}{images/}{./}} %

\usepackage{times}                     %

\usepackage{tasks}

\usepackage{tabu}                      %
\usepackage{booktabs}                  %
\usepackage{lipsum}                    %
\usepackage{mwe}                       %

\usepackage[T1]{fontenc}

\usepackage{mathptmx}                  %

\usepackage{xcolor}

\newcommand{\projectName}{GQVis}
\newcommand{\codeangle}[1]{\texttt{\textless #1\textgreater}}
\newcommand{\code}[1]{\texttt{#1}}
\newcommand{\codebold}[1]{\texttt{\textbf{#1}}}

\onlineid{0}

\vgtccategory{Research}

\vgtcinsertpkg

\title{GQVis: A Dataset of Genomics Data Questions and \\ Visualizations for Generative AI
}

\author{
Skylar Sargent Walters\\
    \parbox{1.4in}{\scriptsize \centering Brown University \\ Harvard Medical School}
\and Arthea Valderrama\\
    \parbox{1.4in}{\scriptsize \centering University of Massachusetts, Lowell \\ Harvard Medical School}\\
\and Thomas C. Smits\\
    \scriptsize Harvard Medical School
\and David Kouřil\\
    \scriptsize Harvard Medical School
\and Huyen N. Nguyen\\
    \scriptsize Harvard Medical School
\and Sehi L'Yi\\
    \scriptsize Harvard Medical School
\and Devin Lange\\
    \scriptsize Harvard Medical School
\and Nils Gehlenborg\thanks{e-mail: nils@hms.harvard.edu}\\
    \scriptsize Harvard Medical School}

\teaser{
  \centering
  \includegraphics[width=\linewidth]{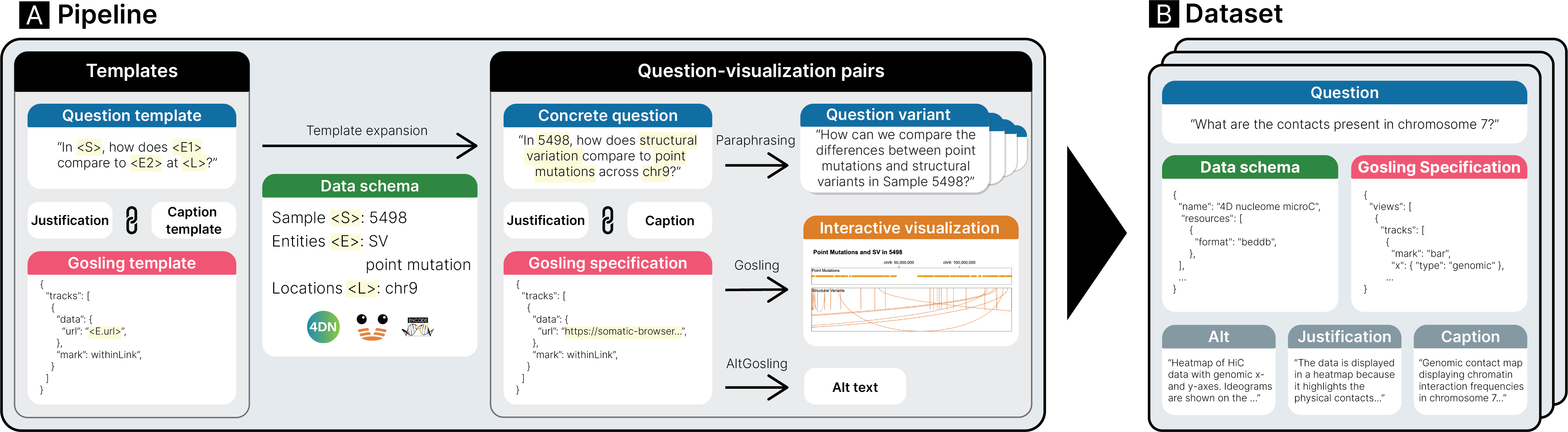}
  \caption{GQVis pipeline and dataset. A) Question-visualization pairs are formed by expanding templates with concrete data sources. 
  B) Each dataset entry consists of a natural language query paired with five response components: a data schema, a Gosling~\cite{lyi2021gosling} specification (renderable as interactive visualization), alt-text, a justification for the response and caption for the figure.
  }
  \label{fig:teaser}
}

\abstract{
Data visualization is a fundamental tool in genomics research, enabling the exploration, interpretation, and communication of complex genomic features. While machine learning models show promise for transforming data into insightful visualizations, current models lack the training foundation for domain-specific tasks.
In an effort to provide a foundational resource for genomics-focused model training, we present a framework for generating a dataset that pairs abstract, low-level questions about genomics data with corresponding visualizations.
Building on prior work with statistical plots, our approach adapts to the complexity of genomics data and the specialized representations used to depict them. We further incorporate multiple linked queries and visualizations, along with justifications for design choices, figure captions, and image alt-texts for each item in the dataset.
We use genomics data retrieved from three distinct genomics data repositories (4DN, ENCODE, Chromoscope) to produce GQVis: a dataset consisting of 1.14 million single-query data points, 628k query pairs, and 589k query chains. 
The GQVis dataset and generation code are available at \url{https://huggingface.co/datasets/HIDIVE/GQVis} and \url{https://github.com/hms-dbmi/GQVis-Generation}.

} %

\begin{document}

\firstsection{Introduction}
\maketitle

The rapid growth of genomic data has created unprecedented opportunities for discovery \cite{brittain_rise_2017}\cite{bick_genomic_2024}\cite{sadasivan_genomic_2024}. However, extracting meaningful insights from these data often requires navigating specialized interfaces, mastering domain-specific visualization tools, and understanding complex data formats. Furthermore, many existing genomic visualizations are static, limiting researchers' abilities to explore data dynamically or reconfigure views to test new hypotheses. Natural language interfaces (NLIs) can address this challenge by harnessing the flexibility and expressivity of natural language, allowing users to articulate what they want to see while generating visualizations that might be difficult to create through conventional interfaces. This approach enables more intuitive, query-driven exploration of complex genomic datasets.

Generative AI has expanded the potential of NLIs, enabling them to interpret user intent and queries to synthesize tailored genomic visualizations on demand. Such capabilities promise to transform how researchers interact with genomic data, lowering the barrier to advanced analysis and accelerating scientific discovery. Beyond immediate visualization, these systems can support downstream applications, such as visual quality assessment, figure captioning, and accessible alternative text generation, broadening the impact of visualization research in genomics.

However, generative NLIs require large and diverse training databases. These typically consist of natural language queries about relevant data and corresponding visualizations. Although general-purpose and biomedical-focused natural language to visualization (NL2VIS) datasets exist \cite{lange_dqvis_2025, ko_vega-lite_nodate, luo_nvbench_2021}, they do not capture the complexity and domain-specific terminology of genomic data. These data are diverse in both file types (e.g., BigWig, VCF, BAM/SAM) and visualization methods (e.g., circular vs. linear layouts, sashimi plots, connectivity plots), and these resulting visuals may not be interactive. Furthermore, there is a disconnect between visualization research and genomic research \cite{pandey_genorec_2023}. This disconnect means that visualization innovations often fail to address the specialized multi-scale navigation and domain-specific terminology required for genomics, while genomic tools may not benefit from the advances in visualization research. Without training data rooted in these visualization conventions and the connection between these fields, even advanced NLIs cannot produce meaningful genomic visualizations. Thus, a dedicated dataset of genomic-specific NL2VIS data is essential to enable NLIs to support exploratory analysis.

To provide the foundation for generative NLIs in genomics, we present \projectName, the first comprehensive dataset of natural language to visualization data for genomic applications (\autoref{fig:teaser}). Our \textbf{primary contribution is a dataset of over 2.2 million data points}, which consists of 1.14 million single-queries, 628k query pairs, and 589k multi-step (3+) query chains, each built using Gosling's interactive visualization grammar \cite{lyi2021gosling}. Our \textbf{secondary contribution is extending the DQVis pipeline} for greater robustness in applied domain settings \cite{lange_dqvis_2025}. The original framework packages data schema, natural language queries, and visualization specifications. We extend this framework to support genomic data formats and incorporate three additional components: 1) justifications describing the choices made in visualization design, 2) academic figure captions, and 3) AltGosling \cite{smits2024altgosling} alternate text descriptions for visual accessibility. These additions increase the granularity and applicability of the dataset, providing rich semantic context for visualizations and supplementary information to enable LLMs to reason more effectively about visual design choices.

\section{Background \& Related Work} 

Over the past two decades, advances in high-throughput sequencing technologies have generated an unprecedented volume of genomic data \cite{reon_biological_2016}. Visualization offers a means for exploration, discovery, and communication within this data, transforming complex numerical representations into interpretable images. Researchers have developed numerous visualization systems to facilitate this analysis of genomic data. These include IGV \cite{thorvaldsdottir_integrative_2013}, UCSC Genome Browser \cite{karolchik_ucsc_2009}, Ensembl \cite{harrison2024ensembl}, and Comparative Genome Viewer \cite{rangwala_ncbi_2024}, each designed to address specific visual and analytical requirements. More recent work that provides multiscale navigation and interaction is Gosling, a grammar-based visualization library for genome track data \cite{lyi2021gosling}. Unlike static visualization grammars, Gosling supports fully interactive, declarative specifications, enabling users to dynamically adjust scales, filter data, and compose multiple coordinated views in real time. This flexibility allows researchers to reproduce established plot types and iteratively explore new hypotheses by reconfiguring and interacting with visualizations.

Outside of genomics, grammar-based visualization systems have been primarily developed in the context of general-purpose data visualization. For example, Vega-Lite is a declarative grammar that specifies visualizations in a JSON format \cite{satyanarayan_vega-lite_nodate}. By abstracting visual design into composable building blocks, Vega-Lite enables reproducibility, flexibility, and diverse visualization types. These general-purpose systems do not support unique visualization methods and file formats commonly used in genomics. In general, grammar-based systems are particularly useful for constructing NL2VIS datasets because they enable structured and constructive specifications, creating alignment between queries and their corresponding visualization components.

Despite great potential in natural language-driven visualization generation, there are no existing datasets for NL2VIS that apply to genomic data. Rather, prior work has demonstrated the feasibility of creating large data visualization repositories for general-purpose and biomedical data. For example, Ko et al. introduced a dataset built on the Vega-Lite grammar \cite{ko_vega-lite_nodate}, which contains a wide range of visualization specifications paired with natural language queries. This method scraped Vega-Lite specifications, then employed an LLM to produce possible queries that would result in the image. Similarly, Luo, Tang, and Yi put forth nvBench, which transformed existing data that mapped natural language queries to SQL queries and instead generated grammar-based visuals \cite{luo_nvbench_2021}.

More recently, the DQVis framework \cite{lange_dqvis_2025} proposed a systematic approach for generating such datasets by varying data attributes and chart configurations. This pipeline creates triples of data, queries, and visualizations, which can be adapted to any grammar-based language of visualization. Our work acts as a proof of concept for extending the DQVis generation pipeline by adapting its use to build a genomic dataset. This implementation demonstrates that grammar-based methods of visualization, when combined with DQVis queries, data schema, and specifications, can produce diverse and meaningful NL2VIS datasets. 

\section{Dataset Generation} 

The dataset generation process consists of five major components: template generation, template expansion, multi-step query curation, paraphrasing, and quality review.

\subsection{Template Generation}
The goal of this step is to capture a range of queries that could possibly be posed for a dataset. Nusrat et al. \cite{nusrat_tasks_2019} details seven abstract tasks ``covering the most important tasks for genomic visualizations.'' We designed abstract queries spanning these seven tasks across single- and multi-locus objectives and single- and multi-feature sets.

In DQVis, abstract queries are written with entity (\texttt{\textbf{E}}) and field (\texttt{\textbf{F}}) placeholders, in which entities refer to donors, patients, or data tables, while features correspond to attributes of an entity. However, this cannot capture the full complexity in the structure of genomic data. Consequently, we expand this placeholder vocabulary to include:
\begin{enumerate}
    \item Sample (\codebold{S}): A given sample or donor. This can contain metadata attributes, such as cancer type, cell type, and tissue type.
    \item Entity (\codebold{E}): A data type found in a sample, such as point mutation data, RNA-seq reads, or Hi-C data. 
    \item Locus (\codebold{L}): A physical location of a gene or genetic marker on a chromosome. 
\end{enumerate}

These placeholders allow us to create generalized query forms that can later be expanded to a number of diverse outputs. For example, instead of writing a dataset-specific query, ``What structural variants are present on chromosome 1?'', the template query would ask ``What  \codeangle{E} are present on  \codeangle{L}?''. Here,  \codeangle{E} will take the place of an entity, while  \codeangle{L}  will take the place of a location. 

Queries may include information on metadata-level, as indicated with \code{S.metadata-identifier} syntax. To represent a cancer type or a cell type, a query would use \code{S.cancer-type} or \code{S.cell-type}, respectively. These queries are paired with a Gosling specification that includes the same placeholders. For example, where the specification asks for a URL, the template will state \codeangle{E.url}, indicating that the place will later be filled with the URL from the corresponding entity. 

For each query-visualization pair, we also create a justification and a caption to provide additional context. The justification describes why the visualization was selected for the query and can include descriptions supporting the use of a circular or linear layout, view alignment, choice of plot type, and more. The caption represents a figure caption for the image, and similar to the queries and visuals, contains \code{S}, \code{E}, and \code{L} placeholders to be expanded for caption specificity.

\subsection{Template Expansion}

The goal of template expansion is to fill the placeholder values in each query and specification with concrete sample, entity, and location names. However, not all abstract queries are logically meaningful. For example, we cannot ask about creating a point plot of structural variants. Therefore, each query-visualization pair will contain constraints that limit which samples, entities, and locations may be applied to the query. If we are asking a question with a corresponding bar graph as the output visualization, we would add the constraint that the \code{E} must be able to be visualized as a bar graph. These constraints can also apply to relationships. If we are asking a question about \code{S1} versus \code{S2}, we must ensure that \code{S1} is not the same as \code{S2}. These constraints ensure a valid output query is created.

After constraints are defined, abstract queries will then be filled with real data. These data come from schemas that describe the sample-level, entity-level, and location-level data for the corresponding datasets. For our schemas, we drew from 4DN \cite{dekker20174dn}, ENCODE \cite{encode2012integrated}, and Chromoscope \cite{lyi_chromoscope_2023} to represent genomic data across structural, functional, and epigenetic applications. 

The reification of abstract queries with a dataset schema is formulated as a constraint satisfaction problem. This locates all sample-entity-locus combinations within the dataset that meet the required constraints and provides a list of solutions that map the abstract features to real data features. Any placeholder names and data references will then be replaced with the concrete data, resulting in a meaningful query and corresponding Gosling specification, which can be converted into an interactive visualization. We also create alternative text from this specification with AltGosling~\cite{smits2024altgosling}.

\subsection{Multi-step Query Curation}

Multi-step query chains are sets of two to eight queries that represent a synthetic analysis sequence. For example, the first query may ask to show the data at SMAD4, while the follow-up could ask to compare this plot to the data at BRCA1. These chains can help train conversational models to update figures in accordance with user requests. 

We first took all major start queries (i.e., the output of single-query template generation) and created a list of possible follow-up queries for each. These possible follow-ups fell into one of five transition types:

\begin{itemize}
    \itemsep 0.5em 
    \item \textbf{Layout}: altering the visual layout of a plot
    \item \textbf{Comparative addition}: stacking a new data view to compare with the initial visual
    \item \textbf{Overlay}: overlaying two visualizations 
    \item \textbf{Location zoom}: focusing the visual on a new location 
    \item \textbf{Data stratification}: stratifying data by type, such as by type of structural variant
\end{itemize}

We tracked chains as tuples of the start query, the follow-up query, and the transition type. The chain adopts a structure related to a linked list, wherein each specification inherits its initial visualization from the previous query and links forward as the next tuple's starting query until the end of the chain. This terminates after a randomly-selected length has been reached, chaining 2--8 queries. 

\begin{figure}
    \centering
    \includegraphics[width=1\linewidth]{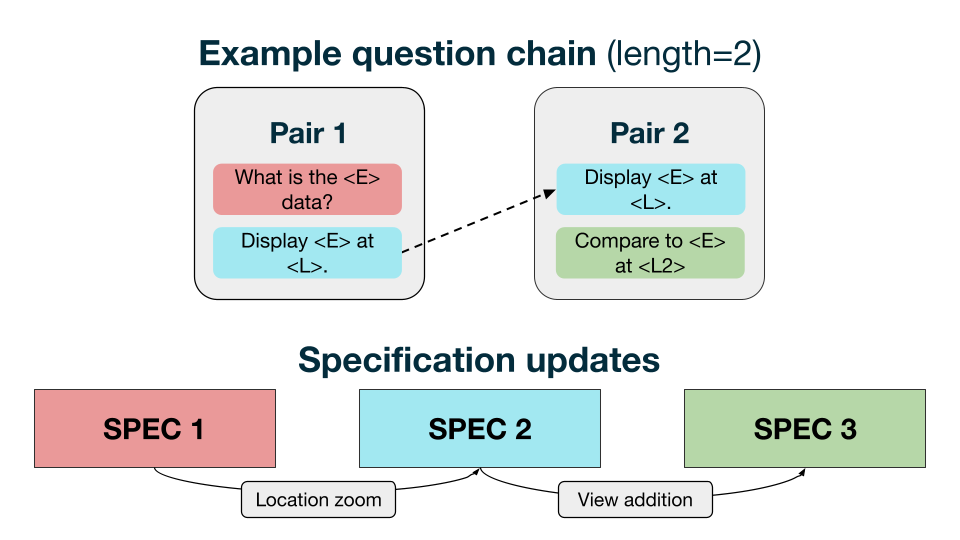}
    \caption{Multi-step generation pipeline. Chains are constructed from pairs of queries, in which the end query of a given pair corresponds to the start query of the next pair. The type of transition will determine how to update specs within the chain.}
    \label{fig:multistep}
    \vspace{-5mm}
\end{figure}

After these multi-step chains are generated, we create concrete queries and visualizations for each query step. Based on the given transition type between two queries, we adjust the output specification to match the new view. For example, suppose we have the initial query ``What is the \codeangle{E} data?'' and the follow-up ``Display \codeangle{E} at \codeangle{L}.'' This type of transition is a location zoom, as we are changing the initial data view from covering the whole genome to covering the area around \codeangle{L}. As a result, the specification will be adjusted according to the handling of a location zoom, creating a new visualization built on the context of the prior query (\autoref{fig:multistep}). 

\subsection{Paraphrasing}

Expanding query templates results in many queries of the same template format. However, these expanded templates do not capture the full diversity and syntax of real user queries. Paraphrasing these concrete queries creates diversity in the query base, adding greater expressivity to the queries. Furthermore, integrating a range of query syntax enriches potential LLM learning from the dataset. 

Based on the Ko et al.~\cite{ko_natural_2024} framework, we employ GPT-4o to vary a query by \textit{expertise} and \textit{formality} on a score of 1--5, with higher scores expressing more technical and proper verbality. Up to 25 paraphrased queries can be generated for each expanded template. Within the prompt template for the LLM, we also input relevant information about a query's dataset schema, such as the entity and sample names, to enhance the LLM's contextual understanding. Thus, a query phrased as ``What is the frequency of structural variants at FBXW7?'' is reworded as ``What is the prevalence of structural variants (SVs) at the FBXW7 location?'' and ``How common are structural variants (SVs) around FBXW7?'', all corresponding to the same visualization.

\subsection{Quality Review Software}

\begin{figure}[h]
    \centering
    \includegraphics[width=0.85\linewidth]{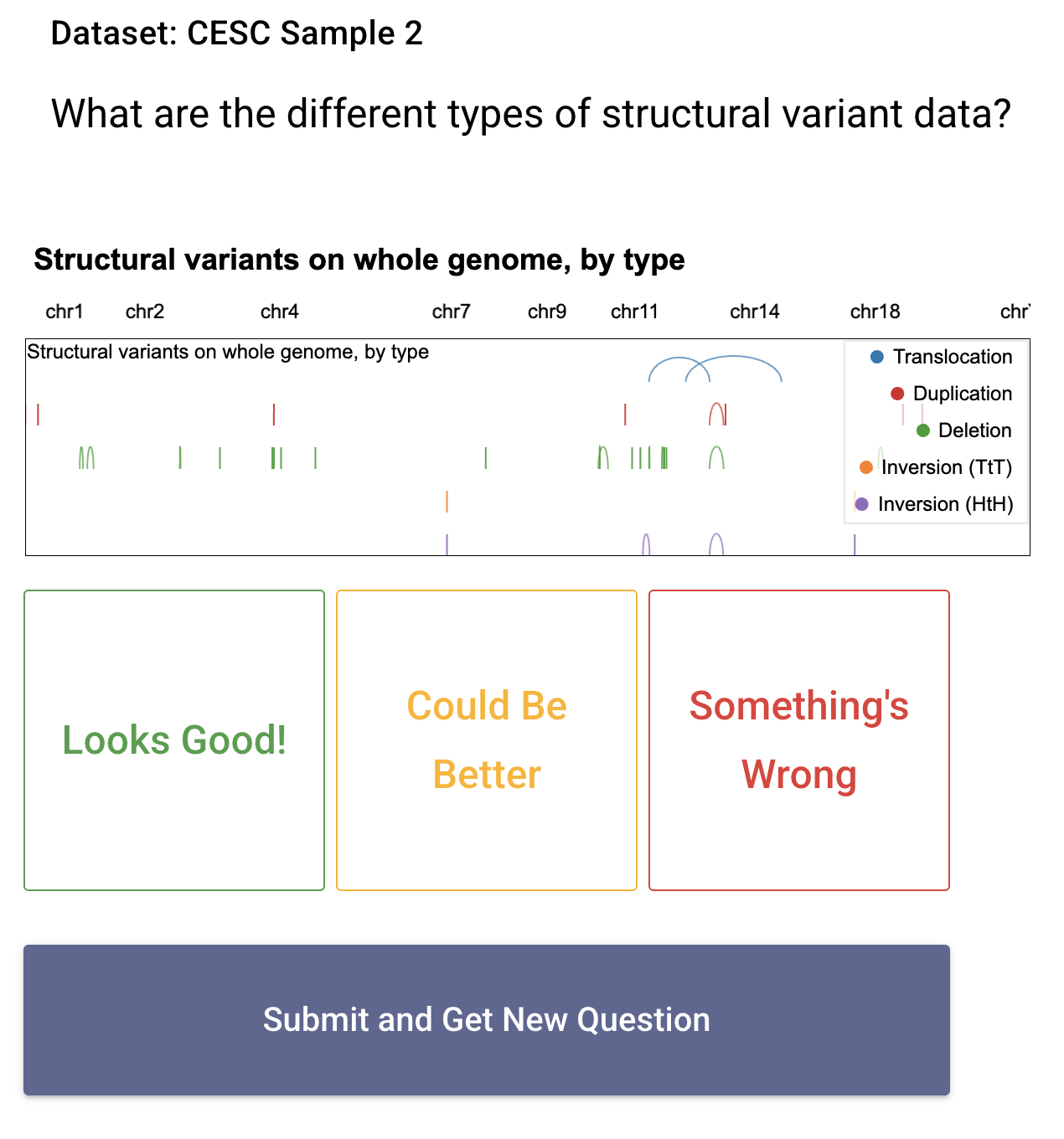}
    \caption{The review interface for assessing the quality of \projectName. Visuals and queries can be reviewed.}
    \label{fig:review}
\end{figure}

The goal of reviewing is to ensure the quality and robustness of the generated dataset. Our review software, shown in 
(\autoref{fig:review}), demonstrates queries and their corresponding Gosling visualizations with options for feedback. Should a given datum be below standards, researchers will have the option to elaborate on the issue and its significance. We plan to obtain opinions from domain-expert scientists across genomic research to assess the datasets to ensure alignment with researchers' goals. Incorporating a review phase enables us to have a dataset that is both applicable to researchers' needs and of high quality. 

\section{Dataset Results} 

\begin{figure}[h]
    \centering
    \includegraphics[width=1\linewidth]{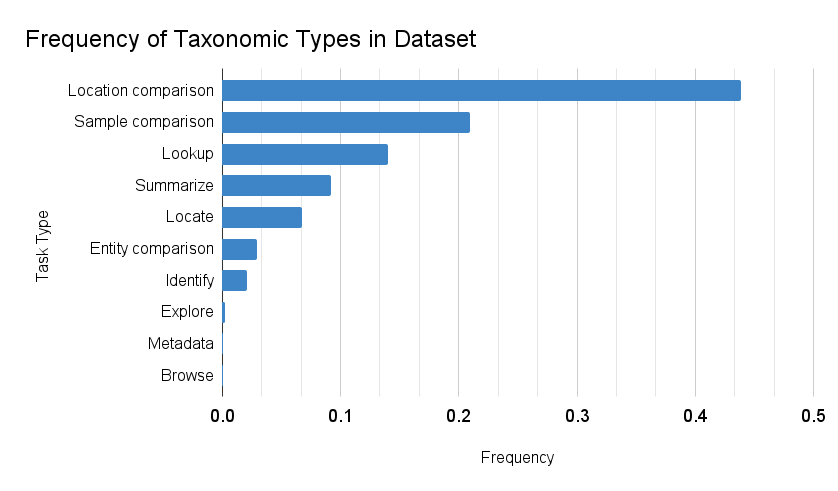}
    \caption{Relative frequency of query-visualization pairs by task taxonomy \cite{nusrat_tasks_2019} after subsampling.}
    \label{fig:taxonomy}
    \vspace{-5mm}
\end{figure}

The initial resulting dataset consisted of 2.2 million data points describing genomic NL2VIS data. However, these data were strongly skewed to represent sample comparison queries, which covered over 80\% of the dataset. To mitigate the bias for specific query types, we implemented data balancing measures to subsample from sample comparison and location comparison queries (\autoref{fig:taxonomy}). Thus, the resulting single-query dataset consists of 1.14 million data points. Subsampling reduced the relative proportion of comparison queries, though they remain the dominant task types. The dataset is therefore not fully balanced, but the adjustment ensures improved representation of other query types without removing the natural distributional skew present in real-world tasks.

\begin{figure}[h]
    \centering
    \includegraphics[width=1\linewidth]{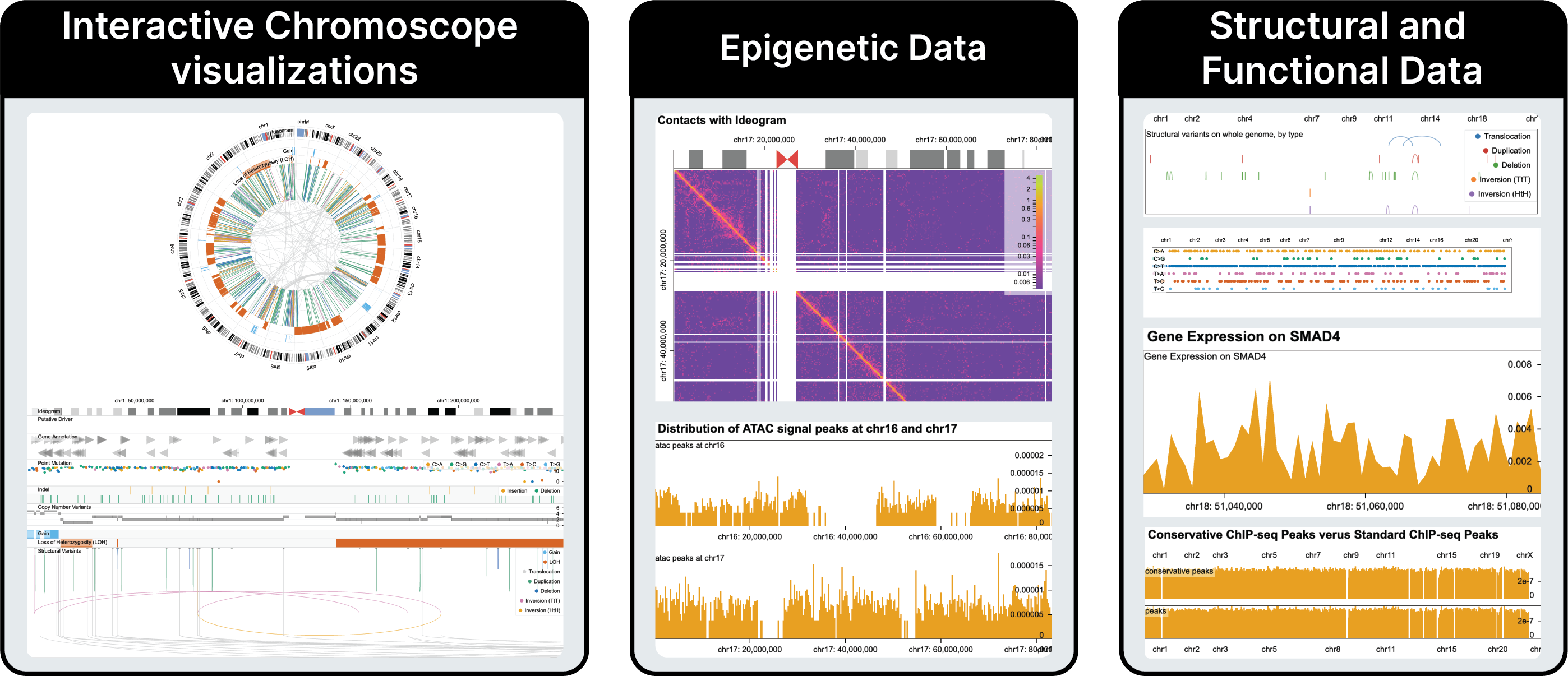}
    \caption{Diversity of the dataset. GQVis covers structural, functional, and epigenomic data, including Chromoscope visuals.}
    \label{fig:diversity}
    \vspace{-5mm}
\end{figure}

The dataset has extensive coverage and diversity of visualizations (\autoref{fig:diversity}). We can view standard \textbf{structural} and \textbf{mutation} data through point, bar, and connectivity plots. Furthermore, epigenetic signals, such as \textbf{Hi-C}, \textbf{ATAC-seq}, and \textbf{ChIP-seq}, can be shown as a range of heatmaps, line plots, bar plots, and area plots depending on the use case. These visualization types support comparison across entities (i.e., ChIP-seq versus ATAC-seq), samples (SV in sample 1 versus sample 2), and locations (Hi-C at chromosome 1 versus chromosome 2). This greatly increases the applicability of the data to investigative use. Moreover, any visualization can be paired with an ideogram for genome context information.

In addition to these single- and multi-view visualizations, we can also import complex visualizations. Chromoscope is a collection of interactive multiscale visualizations for structural variation in human genomics. Each visual combines structural variants, indels, point mutations, a chromosome cytoband, and additional data to create an all-encompassing site for data exploration. These visualizations are directly ported into the GQVis dataset, enabling the creation of highly complex exploratory visualizations across scales.

\section{Conclusion and Future Work} 

This paper 1) generates a dataset of over 2.2 million data points linking natural language queries to data visualizations for genomic applications and 2) proposes a pipeline for generating a natural-language-to-visualization dataset that focuses specifically on genomic data visualizations. We identify two primary directions for future work.

First, we are actively developing a quality assessment framework for the generated dataset. Following the methodology established by DQVis, we are implementing a review interface that enables human evaluators to systematically assess the alignment between generated visualizations and corresponding natural language queries. Our evaluation approach encompasses both individual assessment at the query-visualization pair level and comprehensive robustness evaluation of the entire dataset, representing a work package that extends beyond the scope of this initial contribution.

Second, we plan to leverage this dataset for fine-tuning large language models specifically for natural-language-to-visualization (NL2VIS) tasks in the genomic domain. In particular, we aim to develop a domain-adapted model to generate accurate visualization specifications that respect genomic conventions. Beyond accuracy, we will examine the ability of these models to generalize to unseen query types, thereby establishing benchmarks for NL2VIS systems in genomics and informing the development of next-generation interactive analysis tools. Furthermore, the broad nature of our dataset beyond the standard query/visual system allows us to integrate the data into other tools, such as in visualization quality assessment, caption generation, or multimodal genomics visualization search engines~\cite{nguyen2025multimodal}.
An example application of the fine-tuned LLM is employing it in a visualization authoring tool for genomic data (e.g., Blace \cite{blace}).
One of the biggest challenges that people in genomics confront when authoring visualizations is configuring coordinated multiple views \cite{van2024understanding}, which is essential for visual exploration and analysis of genomic data \cite{l2022multi}.
With the fine-tuned LLM, visualization authoring tools can provide reusable templates of multi-view visualizations based on user prompts describing visualization needs, or recommend the coordinated interactions between pre-authored views.

In summary, this work establishes the first large-scale, genomic-specific NL2VIS dataset and demonstrates the feasibility of adapting grammar-based pipelines to complex genomic domains. By bridging natural language, visualization grammars, and genomic conventions, our approach provides both a resource and a methodology for advancing generative AI-based natural language interfaces. GQVis not only enables model training and evaluation, but also lays the foundation for more accessible, dynamic, and interpretable genomic analysis. As these systems mature, we anticipate that domain-adapted NLIs will lower barriers to exploratory visualization, accelerate hypothesis generation, and ultimately strengthen the integration of computational and biological research.

\acknowledgments{
We thank the members of the HIDIVE lab for their feedback and guidance throughout the project. This work was enabled by the Dr. Susanne E. Churchill Summer Internship in Biomedical Informatics (SIBMI) and the Human BioMolecular Atlas Program (HuBMAP) Undergraduate Student Internship Program. This study was in part funded by NIH grants R01HG011773 and K99HG013348.
}

\bibliographystyle{abbrv-doi}

\bibliography{references, references_manual}
\end{document}